\documentclass[twocolumn]{aa}  % for a paper on 1 column  

\usepackage{txfonts}
\usepackage{graphicx}
\newcommand{\etal}{{et al. }}         % <-- Authors ...{\etal --> et al. } 

\begin{document}

% ------------------------------------------------------------------------------------
% TITLE - AUTHORS - DATES
% -------------------------------------------------------------------------------------
\title{Pulsational frequencies of the eclipsing $\delta$~Scuti star \object{HD~172189}.
       Results of the STEPHI XIII campaign.}

\author{J. E. S. Costa\inst{1},
	E. Michel\inst{1},
	J. Pe\~na\inst{2},
	O. Creevey\inst{3},
	Z. P. Li\inst{4},
	M. Chevreton\inst{1},
	J. A. Belmonte\inst{3},
	M. Alvarez\inst{2},
	L. Fox Machado\inst{3},
        L. Parrao\inst{2},
	F. P\'erez Hern\'endez\inst{3},
	A. Fern\'andez\inst{1},
	J. R. Fremy\inst{1}, 
	S. Pau\inst{1}, \and
        R. Alonso\inst{3}
        }

\offprints{J.E.S. Costa}

\institute{
          Observatoire de Paris, LESIA, FRE 2461, 92195 Meudon, France
          \and
          Instituto de Astronom\'{\i}a - Universidad Nacional
          Aut\'onoma de M\'exico, Ap.P. 877, Ensenada, BC, Mexico 
          \and
          Instituto de Astrof\'{\i}sica de Canarias, 38200 La Laguna, Tenerife, Spain 
          \and
          National Astronomical Observatories, Chinese Academy of Sciences, Beijing 100012, China 
	  }
                     
\date{Received 00.00.0000 / Accepted 22.02.2007}

% -----------------------------------------------------------------------
% ABSTRACT
% -----------------------------------------------------------------------

\abstract{}{}{}{}{} 

\abstract
  % Context (optional).
  {The eclipsing $\delta$~Scuti star \object{HD~172189} is a probable member 
  of the open cluster \object{IC~4756} and a promising candidate target 
  for  the CoRoT mission.}
  % Aims (mandatory).
  {The detection of pulsation modes is the
  first step in the asteroseismological study of the star. Further,
  the calculation of the orbital parameters of the binary system allows us
  to make a dynamical determination of the mass of the star, which works as an important
  constraint to test and calibrate the asteroseismological models.}
  % Methods (mandatory)
  {We performed a detailed frequency analysis of 210 hours of photometric
  data of \object{HD~172189} obtained from the STEPHI XIII campaign\thanks{
    The HD~172189 reduced light curves are available in the CDS via anonymous
    ftp to cdsarc.u-strasbg.fr (130.79.128.5)
    or via http://cdsweb.u-strasbg.fr/cgi-bin/qcat?J/A+A/.
  }.}
  % Results (mandatory).
  {We have identified six pulsation frequencies
  with a confidence level of $99\%$ and a seventh with 
  a $65\%$ confidence level,
  in the range between $100-300\,\mu$Hz. In addiction, 
  three eclipses were observed during the campaign, allowing us to improve the
  determination of the orbital period of the system.}
  % Conclusions (optional).
  {}


\authorrunning{Costa J.E.S. et al.}
\titlerunning{ HD~172189 --- Results of the STEPHI XIII campaign.}

\maketitle

% -----------------------------------------------------------------------
% 1. INTRODUCTION
% -----------------------------------------------------------------------
\section{Introduction}

\object{HD~172189} (\object{IC~4756~93} = \object{SAO~123754}) is a binary star of visual magnitude
$m_V=8.85$ and spectral type A2
 {($\alpha_{2000}=18^h38^m37.554^s$, $\delta_{2000}=+05^o27'55.34''$)}  in the
galactic cluster IC~4756. Recently, Mart\'{\i}n-Ruiz \etal.  (2005) 
(M-R, hereafter) showed
that \object{HD~172189} is an eclipsing binary star with an orbital period of 
$5.71$ days and with $\delta$~Scuti-type pulsations with
a clear frequency of $19.5974\,\rm{cd}^{-1}$ ($226.8\,\mu$Hz),
and  {evidence of more modes} 
in the range $208-232\,\mu$Hz.

The $\delta$~Scuti variables are stars with masses from 1.5
to 2.5 $M_\odot$, located at the intersection of the lower part
of the classical Cepheid instability strip with the 
 {main sequence}.  {They consist of} main sequence objects,
pre-main sequence ones, and also objects that have evolved off the main sequence,
in the hydrogen shell burning phase.
Most of the $\delta$~Scuti  {stars are multiperiodic}, with frequencies between 
50 and 600 $\mu$Hz. 
The pulsation modes are radial as well as non-radial due to
$\kappa$-mechanism associated  {with the zone where He is partially ionized}.
 {Thus, $\delta$~Scuti stars provide} a good opportunity to apply
 {asteroseismology} and study key mechanisms at work in the main sequence
stage,  {among which are transport} of angular momentum, transport of chemical
species at the edge of convective cores  {(overshooting), and large-scale}
circulations.

A few weeks of observations are usually required to resolve their rich
spectra, often showing close pairs of peaks with a spacing of $\sim 0.5\,\mu$Hz.
Three week-long multisite observational campaigns are regularly organized within
the STEPHI network (Michel \etal 1992; Michel \etal 2000),
trying to minimize the problem of missing data due to the day-night cycle,
which induces strong side lobes of the spectral window
in the Fourier spectrum.

The frequency analysis of \object{HD~172189} shows that it is a $\delta$~Scuti with at 
least six pulsation frequencies. Eclipses that occurred during the
campaign were observed, allowing us to  {refine the period} of the binary
system.

The star \object{HD~172189} was found to be a ''likely member'' of the open 
cluster \object{IC~4756} by Kopff (1943). Herzog \etal (1975) updated the results of
a previous work of Seggewiss (1968), concluded that the probability
of the membership  to the cluster for \object{HD~172189}, based on its
proper motion alone, is $92\%$. Missana \& Missana (1995)
re-studied the cluster \object{IC~4756}, estimating the probability of 
membership of each stars and arrived to a similar result for
the case of \object{HD~172189}: $95\%$ of probability 
of being a member of the cluster.

The interest of pulsating stars in cluster and/or in well characterized
binaries for asteroseismology studies has been stressed for long
and by several authors. 
As mentioned in a recent review by Aerts (2006) (see also Aerts and
Harmanec 2004), beside the perspective 
to study processes specific to binarity, like 
the influence of tidal forces on angular momentum transport or eventual
mass transfer, the possibility to determine precisely global parameters
(like individual masses, radii, chemical composition or age) allows to
reduce significantly the parameters space 
to explore in the necessary stellar modeling work.
The seismic interpretation is thus expected to become more 
constraining in terms of diagnostics on physical processes.

In this spirit, Brown \etal (1994) applied singular value
deconvolution technique to study the relative impact of precision on
eigenfrequencies and 
global parameters in the illustrative case of two solar-like pulsators
in a visual binary. They considered the optimal case of the close 
Alpha~Cen binary. Their conclusions necessarily remained theoretical, since
no oscillations were detected at that time neither in Alpha~Cen~A nor in
Alpha~Cen~B. But this seminal work found a great development in Miglio
and Montalb\'an (2005) where an optimization using
Levenberg-Marquardt algorithm is applied to Alpha~Cen~A+B, considering
all available observables, including oscillation frequencies now
detected in both objects. 
This work confirms that satisfying all the observables with present
physical assumptions is becoming more and more challenging and
ineluctably leads to question the description of key-physical processes
like mixing length theory versus full spectrum theory.   
Miglio and Montalb\'an however conclude that the present seismic data do
not allow to go further and discriminate between different physical
options for the equation of state or element diffusion descriptions.

In the specific case of $\delta$~Scuti stars, Creevey \etal (2006) 
proposed to develop the approach of Brown \etal (1994), 
in the specific case of eclipsing binaries featuring a $\delta$~Scuti. 
Their work is however, so far limited only to non seismic observables.

A few $\delta$~Scuti pulsators in eclipsing binaries have already been
intensively searched for oscillations. For RZ~Cas and AB~Cas 
(Rodr\'{\i}guez \etal 2004a; Rodr\'{\i}guez \etal 2004b), 
these observations only brought one oscillation 
frequency, and they did not lead to a seismic interpretation so far. 
RS~Cha is another example of eclipsing binary featuring pre-main sequence
objects, one of them showing $\delta$~Scuti type oscillations (Alecian \etal 2006). 
Recently, new observations have been organized successfully, showing apparent 
multiperiodicity of both component (Bohm \etal In prep.).

The fact that \object{HD~172189} is a $\delta$~Scuti star and at the same time 
a binary system and a member of the open cluster \object{IC~4756} makes it a very interesting
target for the CoRoT mission.

% -----------------------------------------------------------------------
% 2. THE OBSERVATIONS
% -----------------------------------------------------------------------
\section{The observations}

% TABLE 1: JOURNAL OF OBSERVATIONS. --> File: 5784Tab1.tex
\begin{table*}  
  \begin{center}
    
% =================================================================================
% AA/2006/5784 - Costa J.E.S. et al. 
% 
% File....: 5784Tab1.tex
%
% Contents: Table with the log of the observations of the STEPHI XIII campaign. 
%           The letters XL, ZN and SP in the data set names indicate the site of 
%           observation.
%
% =================================================================================

% =================================================================================
% BEGIN OF THE TABLE
% =================================================================================
 \small
 \begin{tabular} {cccccc}
 \hline \hline
 Data set &      Date     & $T_{begin}$   &   $T_{end}$   &   Number  & Duration \\
 name     &               & $245\,0000.+$&$245\,0000.+$ &     of    & (hours)  \\
          &               &     (HJD)    &    (HJD)     &   points  &          \\
 \hline 
 IC0607ZN &   07-Jun-2004 &   3164.51865 &   3164.70145 &     18111 &     5.05 \\  
 IC0607SP &   07-Jun-2004 &   3164.78305 &   3164.94251 &     11327 &     3.61 \\  

 IC0608ZN &   08-Jun-2004 &   3165.44773 &   3165.70545 &     10067 &     2.97 \\   
 IC0608SP &   08-Jun-2004 &   3165.74305 &   3165.97506 &     19082 &     5.37 \\  

 IC0609XL &   09-Jun-2004 &   3166.19328 &   3166.30368 &      9499 &     2.65 \\   
 IC0609ZN &   09-Jun-2004 &   3166.46177 &   3166.70003 &     20586 &     5.72 \\   
 IC0609SP &   09-Jun-2004 &   3166.72632 &   3166.96818 &     20407 &     5.67 \\  

 IC0610XL &   10-Jun-2004 &   3167.07915 &   3167.30217 &     19267 &     5.35 \\  
 IC0610ZN &   10-Jun-2004 &   3167.43952 &   3167.69994 &     22485 &     6.25 \\   
 IC0610SP &   10-Jun-2004 &   3167.72274 &   3167.97030 &     16098 &     4.50 \\   

 IC0611ZN &   11-Jun-2004 &   3168.44450 &   3168.69808 &     21910 &     6.09 \\   
 IC0611SP &   11-Jun-2004 &   3168.69851 &   3168.97172 &     23592 &     6.56 \\  

 IC0612XL &   12-Jun-2004 &   3169.08350 &   3169.29475 &     16178 &     5.07 \\   
 IC0612ZN &   12-Jun-2004 &   3169.43251 &   3169.69744 &     22884 &     6.36 \\  
 IC0612SP &   12-Jun-2004 &   3169.69386 &   3169.97170 &     23435 &     6.52 \\   

 IC0613ZN &   13-Jun-2004 &   3170.43495 &   3170.70100 &     22986 &     6.39 \\  
 IC0613SP &   13-Jun-2004 &   3170.69817 &   3170.97177 &     23591 &     6.57 \\   

 IC0614ZN &   14-Jun-2004 &   3171.42569 &   3171.69950 &     23605 &     6.57 \\  
 IC0614SP &   14-Jun-2004 &   3171.74160 &   3171.96188 &     18238 &     5.07 \\  

 IC0615ZN &   15-Jun-2004 &   3172.41819 &   3172.69828 &     24191 &     6.72 \\ 
 IC0615SP &   15-Jun-2004 &   3172.69699 &   3172.96155 &     22845 &     6.35 \\ 

 IC0616ZN &   16-Jun-2004 &   3173.41763 &   3173.69907 &     24256 &     6.75 \\  
 IC0616SP &   16-Jun-2004 &   3173.69528 &   3173.95722 &     22574 &     6.29 \\ 

 IC0617ZN &   17-Jun-2004 &   3174.41541 &   3174.69871 &     24476 &     6.80 \\  
 IC0617SP &   17-Jun-2004 &   3174.68134 &   3174.96363 &     24302 &     6.77 \\  

 IC0618ZN &   18-Jun-2004 &   3175.41786 &   3175.69765 &     24172 &     6.72 \\  
 IC0618SP &   18-Jun-2004 &   3175.71757 &   3175.96352 &     20072 &     5.58 \\  

 IC0619ZN &   19-Jun-2004 &   3176.42122 &   3176.69892 &     23990 &     6.67 \\  
 IC0619SP &   19-Jun-2004 &   3176.68105 &   3176.96442 &     24959 &     6.96 \\   

 IC0620SP &   20-Jun-2004 &   3177.67549 &   3177.96542 &     24959 &     6.96 \\   

 IC0621SP &   21-Jun-2004 &   3178.68041 &   3178.96473 &     22922 &     6.40 \\   

 IC0622SP &   22-Jun-2004 &   3179.67059 &   3179.96465 &     24506 &     6.84 \\  

 IC06S3SP &   23-Jun-2004 &   3180.68252 &   3180.96416 &     24180 &     6.76 \\  

 IC0624SP &   24-Jun-2004 &   3181.67754 &   3181.95713 &     23220 &     6.46 \\  

 IC0625SP &   25-Jun-2004 &   3182.67586 &   3182.96803 &     25104 &     7.01 \\ 

 IC0626SP &   26-Jun-2004 &   3183.68674 &   3183.70567 &     23307 &     6.52 \\  

 IC0627SP &   27-Jun-2004 &   3184.67712 &   3184.89731 &     21940 &     6.14 \\  

 IC0628SP &   28-Jun-2004 &   3185.67427 &   3185.96532 &     14789 &     4.34 \\ 
 \hline 
 \end{tabular}
% =================================================================================
% END OF THE TABLE
% =================================================================================

  \caption{Log of the observations of the STEPHI XIII campaign. 
           The letters XL, ZN and SP in the data set names
           indicate the site of observation.} 
  \label{tab_log}
\end{center}
\end{table*}

% FIGURE 1: COVERAGE DIAGRAM. --> File: 5784Fig1.eps
\begin{figure} 
  \centering
  \includegraphics[height=9.0cm]{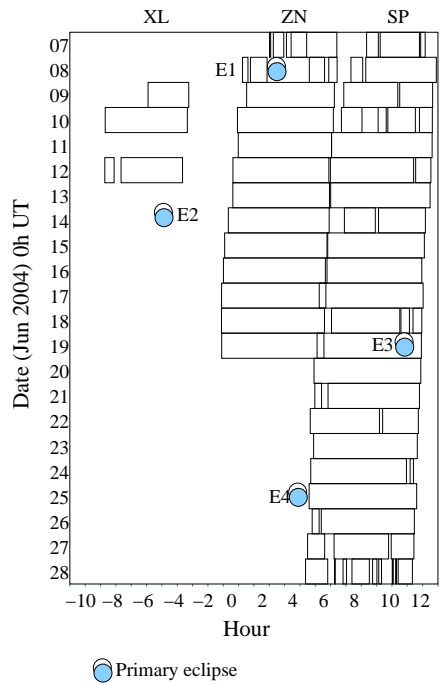}  
  \caption{Coverage diagram of the STEPHI XIII campaign. 
  Each rectangular strip represents a night of
  observation. The labels XL, ZN and SP  {indicate the three observatories}.
  The total number of hours of photometric data
  is 210 with an effective coverage of $40\%$.  {The times of minimum of the} 
  primary eclipses are indicated by E1, E2, E3 and E4.}
  \label{FigCoverage}
\end{figure}

The STEPHI campaigns include three observatories 
around the world: 
Observatorio del Teide (Iza\~na, Tenerife Island, Spain), Observatorio
Astron\'omico Nacional de San Pedro M\'artir (Baja California, Mexico) and
the Xinglong Station of the Beijing Observatory (China). 
The observations were carried out for 21 days over the
period 7-28 June 2004.
The B9 star \object{HD~172248} (\object{IC~4756~117} = \object{SAO 123762}) 
was used as the primary
comparison star ($\rm{comp1}$) with a brightness of $m_V = 8.91$.
Two secondary comparison stars ($\rm{comp2}$) were observed:
\object{IC~4756~142} at the Mexican observatory and \object{IC~4756~115} at the
Spanish observatory. The first one is an A3 star with $m_V = 9.51$ while
the latter is a star of spectral type G0 with $m_V = 10.43$.

We used a four-channel Chevreton photometer at each site. 
Three of the channels are employed to monitor the stars 
(target + comparison 1 + comparison 2), 
and the fourth channel  {is devoted} to  {measuring the sky level}.

 {A total of 210 hours of photometric data  were obtained 
during the  21 days of the campaign.
The integration time was 1 second. This corresponds to a duty cycle
of $40\%$ which is typical for the STEPHI campaigns. }

A log of the campaign  {is given} in Table~\ref{tab_log}. The digits in
the data set names (in the first column) indicate the date  {of the start of the
observations} and the two last letters indicate the site of observation:
XL = Xinglong, ZN = Obs. del Teide (Iza\~na), and SP = San Pedro M\'artir.
Hereafter, we will use these three abbreviations to represent the three sites. 
 {Column 2 gives the UT date at the start of the observations and columns 3 and 4
give the HJD time (Heliocentric Julian Date) at the start and end of the observations.
The last two columns show the number of useful
measurements and the observation length (in hours). }
The coverage diagram of the campaign is shown in Fig.~\ref{FigCoverage}.
The letters in the top of the figure indicate the positions of the
 {observations} from the three sites. The rectangular strips represent the duration
(in hours) of the observations in relation to 0h UT of the date in the vertical axis.
The diagram was made this way to be consistent with Table~\ref{tab_log}.
Note that the observations from each site 
are aligned inside the same approximate range of time. 
 {The times of minimum of the primary eclipses discussed in Sec.~\ref{sec_eclipses}
are indicated in the figure}.

% -----------------------------------------------------------------------
% 3. DATA REDUCTION
% -----------------------------------------------------------------------
\section{Data reduction}

The data  {reduction} followed similar steps as reported in previous 
STEPHI campaigns (see, for instance, Hernandez \etal 1998). 
 {We start with the inspection of the light curve and, when necessary,
elimination of bad points and anomalous parts of the dataset
(including the parts where the eclipses are occurring). }
In the second step we proceed to the calibration
using measurements of the
sky background simultaneously taken  {by all the four} channels, usually
taken at the beginning and at the end of the observations. 
The third step is the subtraction of the sky background.
The measurements of the sky background are subtracted from the
light curve of each channel. However, due to technical problems,
in this campaign  the photometer sky channel could not be used for 
the sky monitoring on some nights.
The fourth step is the division of the light curve of the 
 {variable star} ($\rm{var}$) by the primary comparison star light curve, 
$\rm{var}/\rm{comp1}$,
and by the secondary comparison star light curve, $\rm{var}/\rm{comp2}$.
The purpose of this step 
is to minimize the effect  {of changes 
in the sky transparency on long time scales}.

In order  to  {remove  low-frequency trends} which affect the 
detection of pulsation modes, the next step is to fit 
and  subtract a  {polynomial of low-order} (order $\le 2$) 
from the light curve. 
The last step is to divide the whole light curve by the average value
and subtract $1$. The resulting light curve is the relative
variation in the magnitude of the star  {in relation} to its mean
magnitude.
Most of the light curves show maximum variations around $0.05$ mag,
which corresponds to $\sim 5\%$ relative to the mean intensity.
 {Finally, all times are converted to heliocentric Julian dates (HJD)}.

% -----------------------------------------------------------------------
% 4. FREQUENCY ANALYSIS
% -----------------------------------------------------------------------
\section{Frequency analysis}\label{secAnaFrea}

\begin{figure} 
  \centering
  \includegraphics[width=8.5cm]{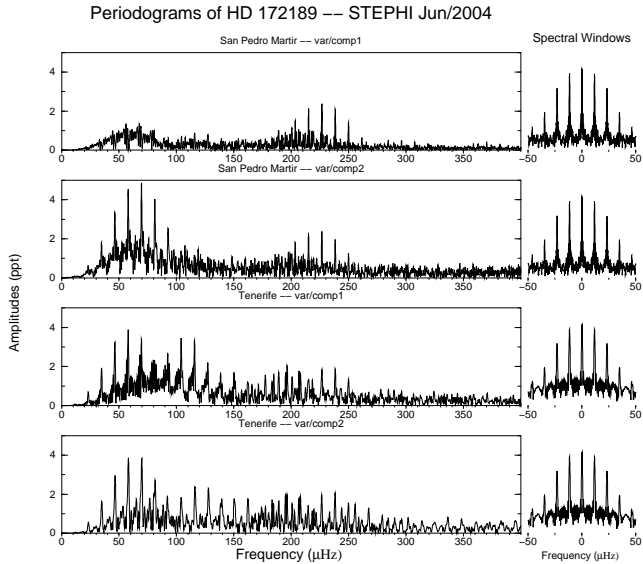}
  \caption{Periodograms of the one-site light curves of 
           $\rm{var}/\rm{comp1}$ and $\rm{var}/\rm{comp2}$ for the
           San Pedro M\'artir and Tenerife data. In the right side
           are shown the respective one-site spectral windows.}
  \label{FigSitesDFTs}
\end{figure}

 {In Fig.~\ref{FigSitesDFTs} we show} the periodograms of the light curves 
of $\rm{var}/\rm{comp1}$ and $\rm{var}/\rm{comp2}$ for San Pedro M\'artir and Tenerife 
data sets for the range of $0-400\,\mu$Hz with amplitudes given in ppt (part per thousand).
The respective spectral windows are shown on the right side.

 {To our surprise}, the periodogram of the light curve of the 
Tenerife data shows the presence of a series of at least 12
harmonics of the frequency of $1\,\rm{cd}^-1$ ($f_d = 11.605\, \mu$Hz).
 {This is not the case} for the San Pedro M\'artir data,
although the reduction procedure is the same for the two data sets.
We discarded the possibility of 
these harmonics being an artifact of an electronic problem in 
one of the channels because they are present in the periodograms of the
light curves of $\rm{var}/\rm{comp1}$, 
$\rm{var}/\rm{comp2}$ and $\rm{comp2}/\rm{comp1}$.
The harmonics, perhaps, are resulting of an occasional change in the
sky transparency at Iza\~na during the campaign.

In the periodogram of the SP light curve $\rm{var}/\rm{comp1}$, we can clearly see
peaks of pulsation modes with high amplitudes and frequencies within the range 
$150-300\,\mu$Hz, completely separated from the bump of peaks of low frequencies 
($f < 150\,\mu$Hz). The same peaks appear in the periodogram of the light curve
of  $\rm{var}/\rm{comp2}$ of the same site, proving that they are related to the
target star and not to the comparison star. 
The difference between the relative heights 
of the bumps in the two cases can be explained by the difference in spectral type: 
the variable star \object{HD~172189} is an A2 star and the first comparison star 
($\rm{comp1}$) is of spectral class B9, 
while the second one ($\rm{comp2}$) is a G0 star.

% FIGURE 2: PERIODOGRAMS: VAR/COMP+VAR/COMP2 OF SP & ZN. --> File: 5784Fig2.eps

% FIGURE 3: PERIODOGRAMS OF ZN DATA WITH/WITHOUT POLYNOMIAL FITTING. --> File: 5784Fig3.eps
\begin{figure} 
  \centering
  \includegraphics[width=8.5cm, height=5.5cm]{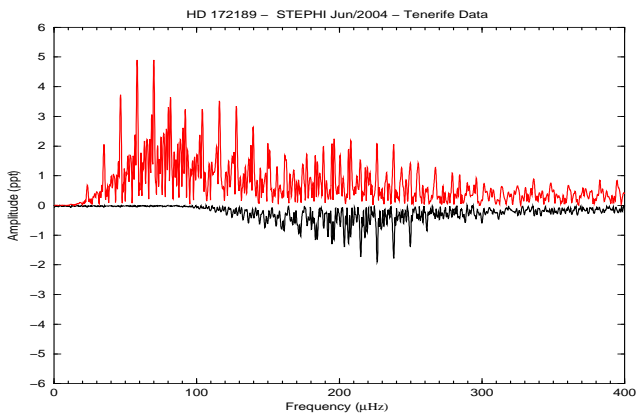}
  \caption{Periodograms of the Tenerife data obtained with polynomial fitting of
           order 2 (top) and order 3 (bottom).}
  \label{FigZN2}
\end{figure}

% FIGURE 4: SPECTRAL WINDOW + POWER OF VAR/COMP2 + POWER OF VAR/COMP. --> File: 5784Fig4.eps
\begin{figure*}
  \centering
  \includegraphics[width=8.0cm]{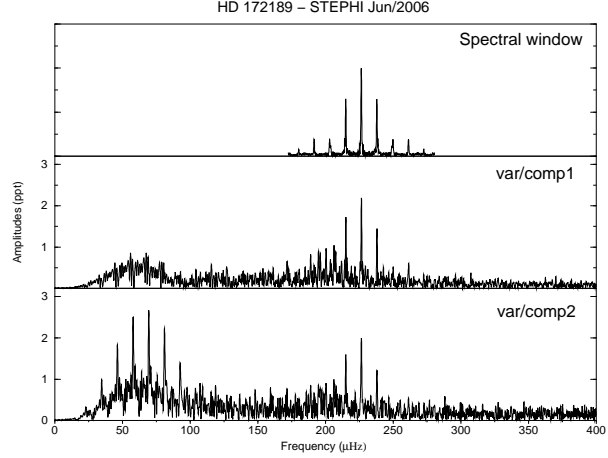}
  \caption{Spectral window (top) and periodograms of the
           whole light curves of HD~172189 of $\rm{var}/\rm{comp1}$ (middle) and
           $\rm{var}/\rm{comp2}$ (bottom).}
  \label{FigPower}
\end{figure*}

The presence of the harmonics complicates the identification of the 
peaks of pulsation frequencies in the spectrum. 
 {We tried a new modification} in the data reduction 
procedure:  we fitted and subtracted a polynomial of order three
(instead of order two) from the Tenerife  {data from each night}.
This procedure was effective
in eliminating all the peaks with high amplitude in the low-frequency
region of the periodograms (0-100$\,\mu$Hz) and the harmonics of the
frequency of the day, 
without any apparent effect over the peaks
with higher amplitude in the pulsation range of frequencies
($100-300\,\mu$Hz) as can be seen in Fig.~\ref{FigZN2}.

In  Fig.~\ref{FigPower} we show the spectral window (top) and the
periodograms of  {the whole light curve} (ZN + SP + XL) of 
$\rm{var/comp1}$ (middle) and $\rm{var/comp2}$ (bottom), 
where we can see
the presence of pulsation modes  {in both cases}.
The central peak of the spectral window 
was arbitrarily positioned
to coincide with the position of the highest peak in the periodogram. 
The FWHM of the central peak is $0.54\,\mu$Hz and 
the two nearest sidelobes have ${\sim}40\%$ of its amplitude. 
All subsequent sidelobes have heights below  ${\sim}4\%$
of the amplitude of the central peak. 
The pulsation frequencies appear in both periodograms, 
in the range $150-250\mu$Hz.

To find pulsation frequencies we used the usual  
iterative approach: 
starting with an empty list of
candidate frequencies and the periodogram of the
original light curve ($\rm{var/comp1}$); 
(1) inside the region of interest in the amplitude spectrum we identify the peaks with
the highest confidence levels (taking care to discard
aliases). If there is no peak with a significant probability
the algorithm stops. 
(2) Put the selected frequencies in the list
of candidate frequencies; 
(3) using a non-linear method,
fit sinusoidals with {\it all} frequencies of the list of
frequencies to the original light curve.
The fitting refines the 
values of the initial frequencies and calculates the amplitudes
and phases, as well as the respective uncertainties. 
(4) The fitted sinusoidals are subtracted from the original
light curve and the  periodogram of the residual
light curve is calculated, and return to step (1)
to search for additional possible pulsation frequencies.

Before applying the algorithm on the light curve $\rm{var/comp2}$, we used a 
high-pass filter to remove signal at low frequencies (< 100 $\\mu$Hz).
As in previous STEPHI articles (e.g. \'Alvarez \etal 1998
Fox-Machado \etal 2002), 
the confidence levels are calculated 
with Fisher's test as prescribed by Nowroozi (1967).
We search for oscillation peaks in the range $100-600\,\mu$Hz,
which in Nowroozi's description 
corresponds to around $m=1000$ ``independent'' frequency
bins. This sets the $99\%$ confidence level at 
$3.4$ times the local square-root of the mean power spectrum
$\sqrt{\bar{A^2_f}}$,
%$\bar{P}^{1/2}_f$, 
which is fitted by the exponential 
$\bar{A^2_f} = c_0\exp(c_1 f)$, where $c_0$ and $c_1$ are constants and $f$
is the frequency.

% - FIGURE 5: PREWHITENING PROCESS. --> File: 5784Fig5.eps
\begin{figure}
  \centering
  \includegraphics[width=8.5cm]{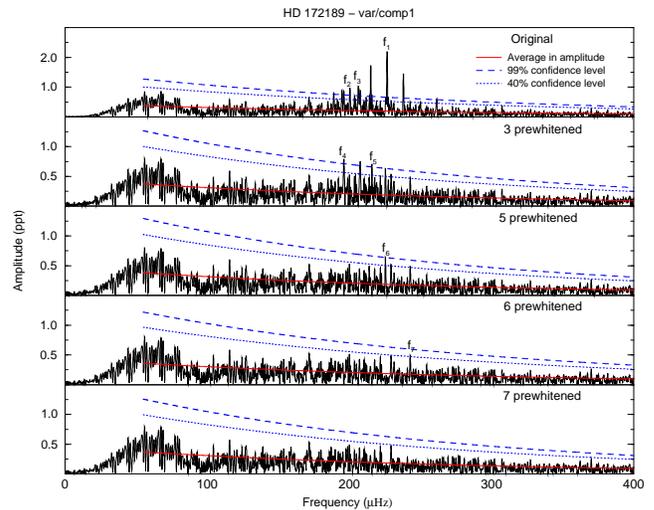}

  \caption{Prewhitening process in HD~172189 ($\rm{var/comp1}$). 
           In each panel the peaks
           above the $99\%$ confidence level (dashed line) are selected
           and removed from the original light curve, together with
           all previous selected frequencies, and a new periodogram
           is obtained. The selected peaks are indicated by $f_1$, $f_2$, etc.
           The lower level of $40\%$ is indicated by the dotted curve and
           the solid curve represents the average amplitude. }
  \label{FigPreWhitening}
\end{figure}

The result of the prewhitening process in HD~172189 
is shown in Fig.~\ref{FigPreWhitening}. 
The dashed curve in each graph indicates the $99\%$ confidence level,
adopted as a detection limit while the dotted curve indicates the position
of the $40\%$ confidence level. 
In the periodogram of the original light curve, we found three peaks above
the $99\%$ confidence level, indicated in the first graph (top) as $f_1$, $f_2$,
and $f_3$. In the two successive prewhitened periodograms we found three more 
frequencies above this level, $f_4$, $f_5$ and $f_6$. After the prewhitening
of the six frequencies, 
only one peak was found with a confidence level upper the 40\% probability limit,
with a 65\% confidence level.
The six detected pulsation frequencies and the seventh frequency with
lower probability are given in Table~\ref{TabFreq}.

We search for the seven detected peaks in the periodogram of the
light curve of $\rm{var/comp2}$ and using a least square fitting we estimated
the probability of each peak not being due to noise in the periodogram of this
light curve: $99.9\%$ for $f_1$; $92\%$ for $f_2$;
$96\%$ for $f_3$; $80\%$ for $f_4$; $30\%$ for $f_5$; $45\%$ for $f_6$; and
$35\%$ for $f_7$. 
As expected, the probabilities found here are lower than those ones found for 
the case of  $\rm{var/comp1}$ because the light curve of $\rm{var/comp2}$ 
is {more noisy}.  {Even so, the four} highest peaks are clearly confirmed in
this test.

M-R found two pulsation modes in \object{HD~172189}:
$226.82\,\mu$Hz and $218.51\,\mu$Hz. The first mode is in agreement with
$f_1$, but the second one does not agree with our frequencies.
However, we identify the peak at 
$218.51\,\mu$Hz as an alias of $f_3$.

% - TABLE 2: 7 DETECTED FREQUENCIES. --> File: 5784Tab2.tex
\begin{table}
  \centering
  \begin{center}
    \begin{tabular} {cccc}
      \hline\hline
      $f_i$ &      Frequency      &  Amplitude        &  Confidence \\
            &     ($\mu$Hz)       &  (ppt)            &  Level      \\
      \hline
      $f_1$ & $226.641 \pm 0.003$ & $  2.19 \pm 0.02$ & $99\%$ \\ 
      $f_2$ & $200.506 \pm 0.006$ & $  0.99 \pm 0.02$ & $99\%$ \\ 
      $f_3$ & $206.270 \pm 0.006$ & $  0.98 \pm 0.02$ & $99\%$ \\ 
      $f_4$ & $196.244 \pm 0.008$ & $  0.81 \pm 0.02$ & $99\%$ \\ 
      $f_5$ & $215.892 \pm 0.010$ & $  0.69 \pm 0.02$ & $99\%$ \\ 
      $f_6$ & $225.446 \pm 0.010$ & $  0.67 \pm 0.02$ & $99\%$ \\ 
      $f_7$ & $242.853 \pm 0.012$ & $  0.50 \pm 0.02$ & $65\%$ \\ 
      \hline
    \end{tabular}
  \end{center}

  \caption{Detected pulsation frequencies in \object{HD~172189} with confidence
            levels $\ge 40\%$.}
          \label{TabFreq}
\end{table}

% -----------------------------------------------------------------------
% 5. SEARCHING FOR ECLIPSES IN HD 172189
% -----------------------------------------------------------------------

\section{Eclipses in HD~172189}
\label{sec_eclipses}

The binarity of \object{HD~172189} was discovered from analysis of
the data of a campaign carried out in 1997 for detecting 
 {$\gamma$~Doradus} pulsating stars in the cluster 
IC~4756 (Mart\'{\i}n 2000; Mart\'{\i}n 2003). 
M-R  {analyzed Str\"omgren 
%$uvb\gamma~\beta$
$uvby-\rm{H}_{\beta}$
} observations obtained
in different campaigns during 1997, 2003, and 2004,
showed the presence of three light minima due to eclipses,
but only one of the eclipses was followed past the minimum
in brightness with a depth of ${\sim}0.12$ mag.
From the data analysis,  {they found  a period of
$P = 5.70198$ days}.
 {The ephemeris obtained from the data is}
$T_{\rm min\,I} = 2\,452\,914.644(3)+5.70198(4)\cdot E$ (in HJD),
where $E$ is the integer number of primary eclipses since the 
epoch $245\,2914.644$ HJD.
The results indicate a system with an orbital eccentricity of
$e\simeq 0.24$, a longitude of the periastron of $\omega \simeq 68^o$
and an inclination of $i\simeq 73^o$. The analysis also suggests
that the two stars have  {different radii} ($r_2/r_1\simeq 0.6$),
but similar temperatures  {$T_{{\rm eff};1}\simeq 1.05\,T_{{\rm eff};2}$}.

According to the ephemeris, there were four primary 
eclipses during the present campaign,
but, as shown in Fig.~\ref{FigCoverage},
only the eclipses $E1$, $E3$ and $E4$
occurred while the star was being observed.
The first one by the Tenerife observatory and
the last two by the Mexican observatory, in the
data sets \verb+IC0608ZN+, \verb+IC0619SP+ and \verb+IC0625SP+, respectively 
(cf. Tab~\ref{tab_log}).

\begin{figure}
  \centering
  \includegraphics[width=8.5cm]{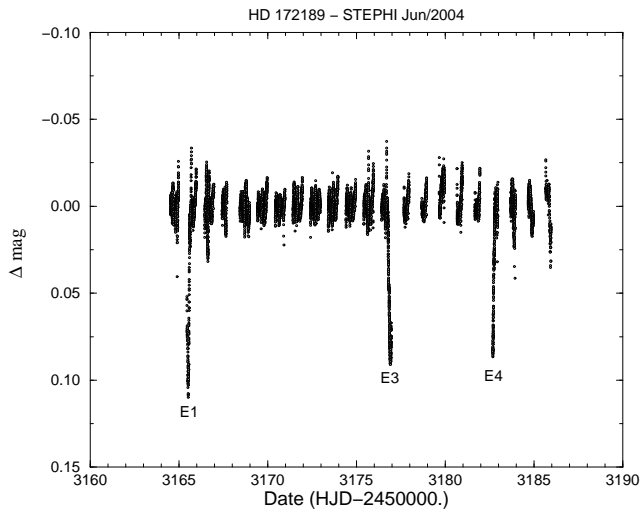}
  \caption{Light curve of \object{HD~172189}. The measurements  {were binned to
           60 seconds}. Three primary eclipses, $E1$, $E3$ and $E4$ are clearly
           visible.}
  \label{FigEclipseLC}
\end{figure}

Fig.~\ref{FigEclipseLC} shows the positions of the eclipses in the
light curve of \object{HD~172189}. 
 {The seven detected pulsation frequencies were subtracted and 
points collected within 60 s were binned}. 
The three primary eclipses are clearly visible.

% FIGURE 6: LIGHT CURVE WITH 3 ECLIPSES. --> File; 5784Fig6.eps

% FIGURE 7: ZOOM - 3 ECLIPSES --> File: 5784Fig7.eps
\begin{figure}
  \centering
  \includegraphics[width=8.5cm]{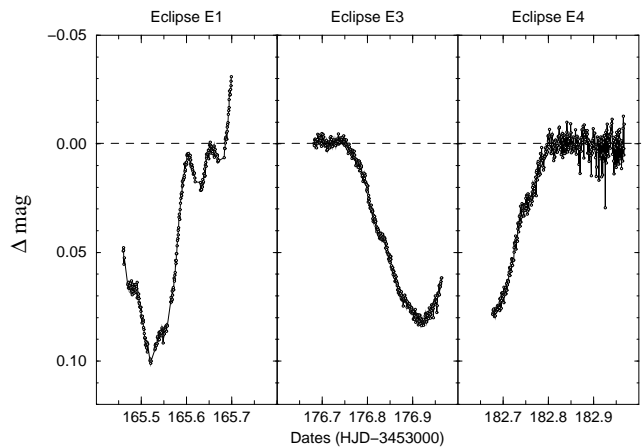}
  \caption{The three primary eclipses ($E1$, $E3$ and $E4$) observed during the campaign. 
           The points were binned and the pulsation frequencies were subtracted.}
  \label{FigEclipse4}
\end{figure}

Fig.~\ref{FigEclipse4} shows the light curves during the eclipse.
 {In the first two panels we  see} the
light decreasing, passing through a minimum in brightness and then increasing. 
The primary eclipse $E1$ has its minimum at $2453165.520$ (HJD).
Unfortunately, the observations were interrupted several times during the
night of this eclipse and the reduction of the data presented some
difficulties.

A better scenario is shown in the second graph of Fig.~\ref{FigEclipse4}
for the primary eclipse  $E3$, where the light curve changes 
slowly and the minimum is well defined.
 {Fitting a polynomial curve of degree 3} to the eclipse part of the light curve, 
we obtain $T_{\rm min\,I} = 2453176.920(2)$ (HJD), which differs by
$\sim 9 \sigma$  {from the predicted value using the ephemeris from M-R}, 
indicating that the uncertainty
in the orbital period is underestimated.
Using this new minimum and the previous ephemeris, 
we recalculated the orbital period to obtain $P = 5.70165(8)$ days. 
Adopting the instant of the primary eclipse $E3$ as the new epoch, 
the new ephemeris is:

\begin{equation}
 T_{\rm min\,I}= 2453176.920(2) + 5.70165(8)\cdot E \quad \mbox{(in HJD).} 
\end{equation}

Table~\ref{TabEclipses} shows the ephemeris dates for the four eclipses.
The calculated date for the eclipse $E1$ is $245\,3165.518(2)$, 
consistent with the minimum observed in the light curve. 
The maximum in depth for the primary eclipses $E1$ and $E3$ are 
 {$\Delta mag=0.10(3)$ and $\Delta mag=0.08(3)$},
respectively, in agreement with the minimum of  {$\Delta mag=0.12$}
for the eclipse observed by M-R.

% TABLE 3: ECLIPSES.
\begin{table}
   \centering
   \begin{tabular}{cc} \hline \hline
   Eclipse    & Date (HJD)        \\ \hline
     E1  & $245\,3165.517(2)$ \\
     E2  & $245\,3171.218(2)$ \\
     E3  & $245\,3176.920(2)$ \\
%%%%%E4  & $245\,3188.323(2)$ \\ \hline     <-- ! WRONG date
     E4  & $245\,3182.612(2)$ \\ \hline     <-- ! RIGHT date
   \end{tabular} 
   \caption{Primary eclipses of HD~172189.}
   \label{TabEclipses}
\end{table}

The light curves show that the duration of each primary eclipse 
is $\sim$ seven hours and the half-depths occur around one hour
before and after the instant of minimum.  
From the orbital elements calculated by M-R, we calculated a phase of
$\sim 0.132$ for the secondary eclipse, i.e., the secondary eclipses would
occur $\sim 0.75$ days ($\sim 18$ hours) after each primary eclipse
and have the same duration. We searched for minima in the light curve around 
these dates, but as was remarked by M-R in their analysis, no minimum with  {an
unequivocal} and statistically significant depth was found.

% -----------------------------------------------------------------------
% 6. CONCLUSIONS
% -----------------------------------------------------------------------
\section{Conclusions}
                                 
The STEPHI XIII campaign carried out in June 2004, allowed a study of the
pulsating behavior of the  {$\delta$~Scuti} star \object{HD~172189}.
{The Fourier analysis of the light curve from three sites shows the presence of
six pulsation frequencies} with a confidence level of 99\%.
An additional frequency with a lower probability of $65\%$ was also found. 
The observation of three eclipses during the campaign --- two of them
including the minimum in intensity ---
allows us to refine the previous value for orbital period of the system,
obtaining  5.70165(8) days.  

% %% [[ --->
% A more detailed study of the binarity of HD172189 aiming
% at the determination of orbital elements and individual 
% masses and radii will require complementary 
% photometric and spectrometric observations.
% This work is under progress.
% As primary target of the CoRoT mission, HD~172189 could be continuously 
% observed for six months, providing a periodogram with very high resolution
% allowing the detection and determination 
% of a larger number of pulsation modes. This would allow us to attempt
% to fit the frequencies to theoretical pulsation models of the star.
% The dynamical determination of the stellar mass works as
% an important constraint to test and calibrate the theoretical models. 
% On other hand, from evolutionary asteroseismological
% models it is in principle possible to calculate
% the age of the star and compare it with
% the estimated age of the cluster IC~4756 using theoretical isochrones.
% % <---]]

With the detection of six pulsation frequency with a high confidence
level, our present campaign confirms the interest of \object{HD~172189} as eclipsing
binary featuring a $\delta$~Scuti pulsator for seismology. 
Besides the number of modes detected, the interest of \object{HD~172189}
Compared with other objects like RZ~Cas (Rodriguez \etal 2004a) or 
AB~Cas (Rodriguez \etal 2004b) is also associated with its orbital period.
Both, AB~Cas and RZ~Cas, reveal a short orbital period ($\sim 1$d). 
This suggests possible important effect of tidal forces
on the star shape and structure, potential effects on oscillations. 
In fact, both these objects are considered Algol-type binaries, 
which supposes that mass transfer occurred between companions and
changed their evolution compared with single stars.  
With a 5.7d orbital period, HD~172189 is expected to be more
representative of a ''normal'' single star, both for its pulsational 
behavior and for its evolution history.

Dedicated spectroscopic and photometric observations are under way to
characterize precisely the orbital parameters and radii of HD~172189 members. 
These results when available, completed by our present oscillation frequencies 
analysis will allow to start the modeling and seismic analysis of this object 
taking advantage of these numerous constraints.
In addition to this, within a few years now, it is planed to observe HD~172189 
with CoRoT for five months. This is expected to allow detection of many more 
oscillation modes by pushing down the noise level by a factor 500
in the power spectrum (see Michel \etal 2006).
Beside this, the fact that \object{HD~172189} belongs to the cluster \object{IC~4756}, even
if this one is not so well studied yet,
offers an interesting potential source of complementary information for
further developments.

% -----------------------------------------------------------------------
% ACKNOWLEDGMENTS
% -----------------------------------------------------------------------

\begin{acknowledgements}
This work received financial support from the Spanish DGES 
(ESP2004-03855-C03-03), the Chinese National Natural 
Science Foundation under grant number 10573023 and 10433010,
and the Brazilian agency - Conselho Nacional de Desenvolvimento 
Cient\'{\i}fico e Tecnol\'ogico, CNPq.
The 1.5 m Carlos S\'anchez Telescope is operated on the island of 
Tenerife by the
Instituto de Astrof\'{\i}sica de Canarias at the Spanish Observatorio 
del Teide.
\end{acknowledgements}

% -----------------------------------------------------------------------
% BIBLIOGRAPHY
% -----------------------------------------------------------------------

%\listofobjects

% -----------------------------------------------------------------------
% END OF THE TEXT BODY
% -----------------------------------------------------------------------

\end{document}